\documentclass{webofc}
\usepackage[varg]{txfonts}   
\begin{document}
\title{Cosmology with galaxy clusters: impact of theoretical and observational systematic uncertainties.}
%
%

\author{\firstname{Laura} \lastname{Salvati}\inst{1,2}\fnsep\thanks{\email{laura.salvati@universite-paris-saclay.fr}} 
}

\institute{Université Paris-Saclay, CNRS,  
    Institut d'Astrophysique Spatiale, 
    91405, Orsay, France
\and
           INAF – Osservatorio Astronomico di Trieste, Via G. B. Tiepolo 11, 34143 Trieste, Italy
          }

\abstract{%
In this talk I focus on how the modelling of the mass-observable relation and the halo mass function can impact the accuracy and precision of cosmological constraints inferred from galaxy clusters.
I present a new analysis of clusters detected in mm wavelengths by the Planck satellite, highlighting the need of an improved description and calibration for the mass-observable relation.
I also discuss how to improve our analysis in view of future cluster surveys, with a particular focus on the impact of the halo mass function calibration and the need for a universal definition.
}
\maketitle

\section{Introduction}\label{intro}
Galaxy clusters are the largest, gravitationally bound structures in the Universe. They are associated with peaks in the matter density field on megaparsec scales. 
The abundance of clusters depends on the underlying cosmological model, tracking the growth of structure and matter density, and therefore it provides constraints on cosmological parameters; see e.g. \cite{Allen:2011zs}. 

In general, when studying galaxy clusters for the cosmological analysis, number counts are used as a cosmological probe. 
Cluster number counts should coincide with the halo mass function, i.e. the number distribution of clusters in bins of redshift and mass. However, cluster masses cannot be measured directly.
Thus, we rely on observables that act as mass-proxies and that tightly correlate with the underlying cluster mass, via some statistical scaling relation. We use these scaling relations, with a model for the cluster selection process, to transform the theoretical mass function into a prediction for the distribution of clusters in the space of survey observables.
The calibration of the scaling relations represents the current limiting systematic in cluster cosmology studies and it is usually referred to as the ``mass-calibration problem"

In this analysis, we consider at first clusters detected in the mm-wavelengths, through the thermal Sunyaev-Zeldovich (tSZ) effect \cite{Sunyaev:1970eu}, by the Planck satellite \cite{Planck:2015koh,Planck:2015lwi}.
We then consider clusters observed in the optical and near-infrared wavelengths, performing a forecasts analysis for future experiments.
We focus on how different modelling and assumptions for scaling relations and mass function can largely impact the constraints on cosmological parameters inferred from cluster number counts. We also comment on how these differences affect the possible tension with other cosmological probe, e.g. the $\sigma_8$ discrepancy between tSZ clusters and cosmic microwave background (CMB) primary anisotropies.

\section{Mass-observable relation calibration}\label{sec-1}

Planck cluster catalog for the cosmological analysis is composed by 439 clusters, in the redshift range $z=[0,1]$ 
and above signal-to-noise ratio $q_{\text{min}} = 6$. For the full description of the number counts modelling, we refer the reader to \cite{Planck:2015lwi}. We recall here the model for the formulation of the scaling relations: under the assumption of self-similarity and hydrostatic equilibrium (HE), we find a relation between tSZ observable and cluster mass, estimated from the tSZ effect, $M_{\text{SZ}}$. However, since this evaluation is based on HE, it provides biased-low masses. In order to take into account this effect, we add in the analysis the mass bias parameter, defined as the ratio between the tSZ mass and the real cluster mass (with the cluster being defined as an over-density of 500 times the critical density of the Universe), $(1-b) = M_{\text{SZ}}/M_{500}$. Recent hydrodynamical simulations provide the estimation $(1-b) \sim 0.8$, see e.g. collection of results in \cite{Salvati:2017rsn,Gianfagna:2020que}.

The calibration of Planck scaling relations is based on a multi-wavelength analysis, using X-ray observations and weak lensing (WL) measurements. In particular, the mass bias calibration is obtained from the CCCP WL analysis \cite{Hoekstra:2015gda}, $(1-b) = 0.780 \pm 0.092$. As discussed in \cite{Planck:2015lwi}, a possible way to reconcile cosmological constraints from clusters and CMB anisotropies points towards a low value of the mass bias, $(1-b) = 0.58 \pm 0.04$. However, this value is not in agreement with results from hydrodynamical simulations, WL evaluations and measurements of cluster gas fraction, as discussed e.g. in the X-COP analysis \cite{Eckert:2018mlz}.

\subsection{Combining tSZ observables}\label{sec:comb_tSZ}
In order to fully understand the CMB-cluster tension and the impact of the mass calibration, we perform a new cosmological analysis of the Planck tSZ signal: we combine the cluster counts ($\text{NC}^{\text{tSZ}}$) with the power spectrum of the full tSZ signal \cite{Planck:2015vgm}, $C_{\ell}^{\text{tSZ}}$, for which we consider also detections from the South Pole Telescope (SPT) \cite{George:2014oba}. The tSZ power spectrum is modelled with the same ingredients as number counts, in particular the scaling relations and the mass function. Combining these two probes would ideally break degeneracy between cosmological and scaling relation parameters, since we find: $\text{NC}^{\text{tSZ}} \propto \sigma_8^9 \Omega_m ^3 (1-b)^{3.6}$ and $C_{\ell}^{\text{tSZ}} \propto \sigma_8^{8.1} \Omega_m ^{3.2} (1-b)^{3.2}$. We start analysing the $\Lambda$CDM scenario and we then consider the impact of varying the total neutrino masses. The full analysis is presented in \cite{Salvati:2017rsn}.

\begin{figure}
\centering
\sidecaption
\includegraphics[width=4.5cm,clip]{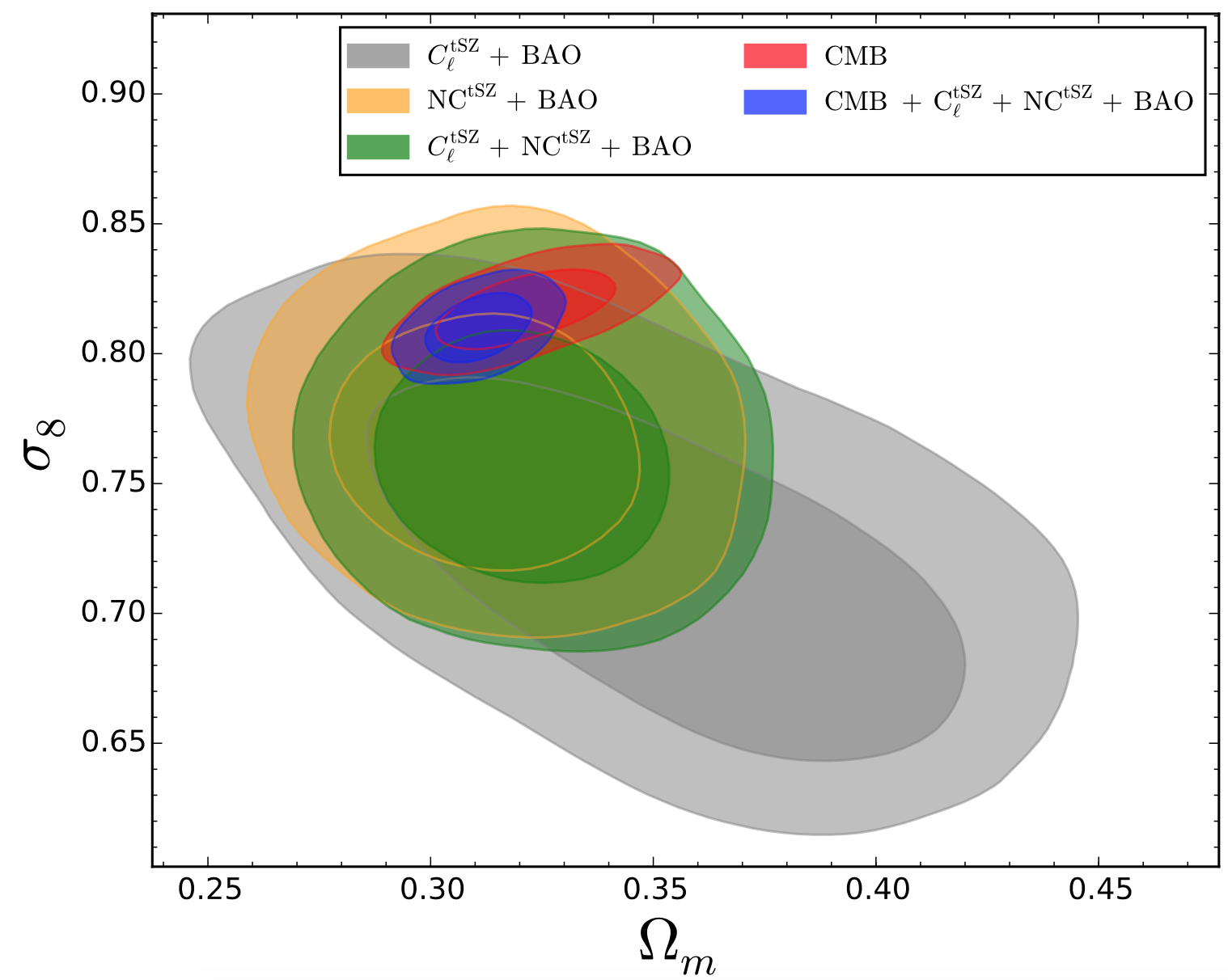} \, \includegraphics[width=4.5cm,clip]{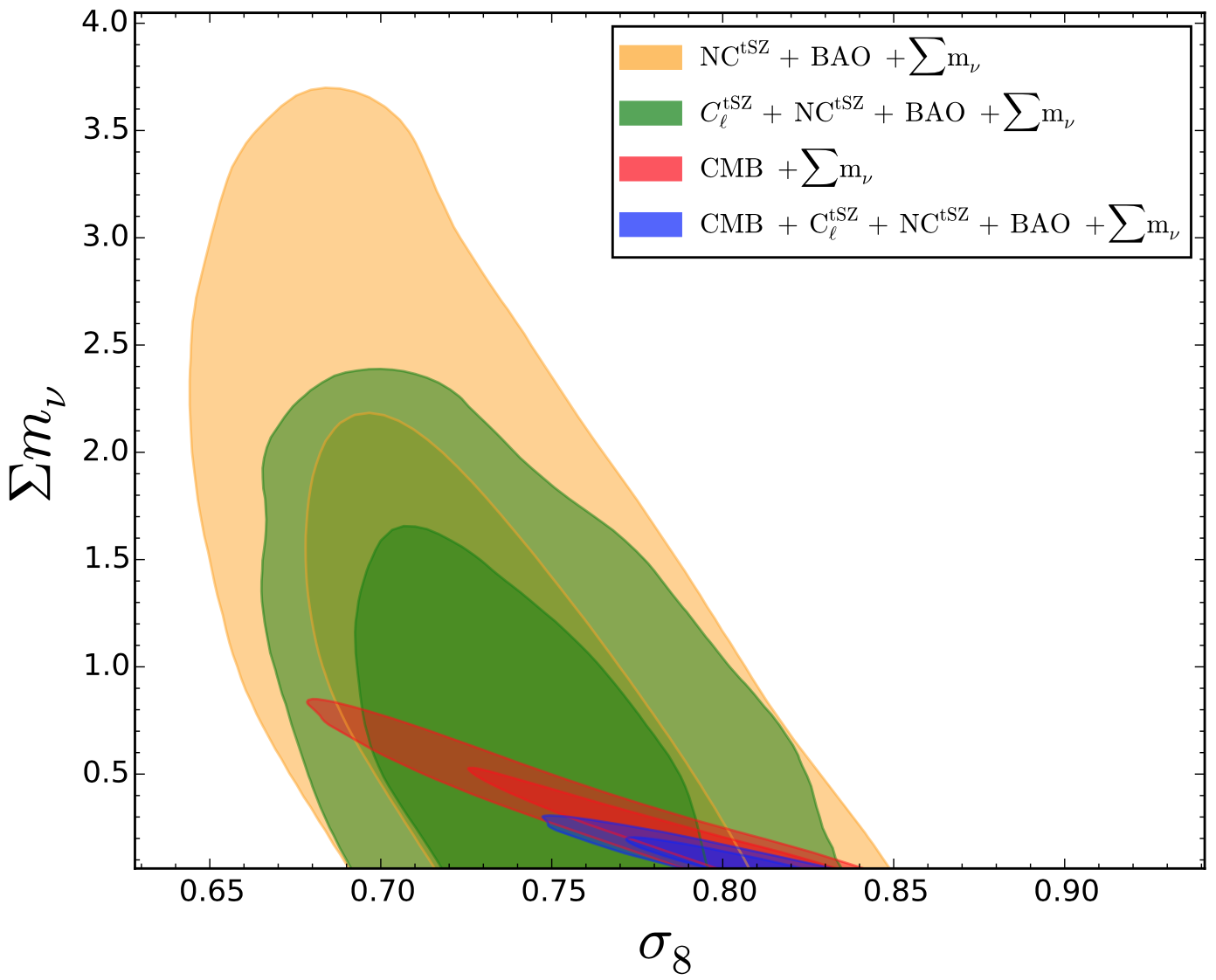}
\caption{Two-dimensional probability distributions for $\Omega_m$, $\sigma_8$ and $\sum m_{\nu}$ for different dataset combinations, as discussed in the text. These plots are taken from \cite{Salvati:2017rsn}.}
\label{fig:p1_1}      
\end{figure}

We show the main results in figure~\ref{fig:p1_1}. On the left panel, we report the two-dimensional probability distributions in the $(\Omega_m,\sigma_8)$ parameter plane for different dataset combinations, when considering the standard cosmological model. First of all, we notice that results for $\text{NC}^{\text{tSZ}}$ + $C_{\ell}^{\text{tSZ}}$ (green curves) combination are dominated by the $\text{NC}^{\text{tSZ}}$  constraining power (orange curves). We then stress that results from tSZ probes and CMB primary anisotropies (red curves) are now fully in agreement, in particular the $\sigma_8$ constraints agree within $2 \, \sigma$.
This is also due to the further improvement in the Planck CMB analysis \cite{Planck:2016kqe,Planck:2016mks,Planck:2018vyg} which now prefers lower values of $\sigma_8$.
In the right panel we analyse results when considering the sum of neutrino masses as a varying parameter. In this case, we find that  $C_{\ell}^{\text{tSZ}}$ contribution strongly improves the constraints, moving the $95\%$ upper limit on massive neutrinos down to $\sum m_{\nu} < 1.88 \, \text{eV}$. 
This is mainly due to the ability of $C_{\ell}^{\text{tSZ}}$ to constrain scales at which massive neutrinos start to have an impact on the evolution of the matter power spectrum (see e.g. \cite{Lesgourgues:2014zoa} for a full discussion of the impact of massive neutrinos on the matter power spectrum). This effect is even stronger when combined with CMB data: we improve the results from $\sum m_{\nu} < 0.68 \, \text{eV}$ (for CMB alone, red curves) to $\sum m_{\nu} < 0.23 \, \text{eV}$ (for CMB+tSZ probes, blue curves).

We now discuss results for the mass bias. We focus on constraints obtained for the CMB+tSZ probes combination, for the standard model of cosmology, $(1-b) = 0.65 \pm 0.04$, and when considering a varying neutrino mass, $(1-b) = 0.67 \pm 0.04$. These constraints are higher than the Planck 2015 analysis \cite{Planck:2015lwi}, due to the shift on the $\sigma_8$ parameter for CMB analysis. Nevertheless, they are still lower than the expected $0.8$ value, pointing to a remaining  inconsistency between CMB primary anisotropies and tSZ probes. We stress that this tension is not solved when exploring extensions to $\Lambda$CDM: we tested also the impact of a varying equation of state (EoS) parameter for dark energy, for which we find $(1-b) = 0.63 \pm 0.04$.

\subsection{Improving the mass bias definition}\label{sec:mass_bias}
In order to further understand this remaining discrepancy on the scaling relation calibration, we analyse a new definition of the mass bias, considering a mass and redshift dependence. We stress that, in general, WL calibrations of the mass bias are based on very small subsamples of the full Planck cosmological cluster sample, considering few tens of objects in a limited redshift range. It is therefore possible that the subsample used for the calibration is not representative of the full cluster sample and a more complex definition for the mass bias is needed. We adopt the following definition:
\begin{equation}\label{eq:bias_var}
(1-b)_{\text{var}} = (1-\mathcal{B}) \left( \dfrac{M}{M_*} \right)^{\alpha_b} \left( \dfrac{1+z}{1+z_*} \right)^{\beta_b} \, .
\end{equation}

\begin{figure}
\centering
\sidecaption
\includegraphics[width=10cm,clip]{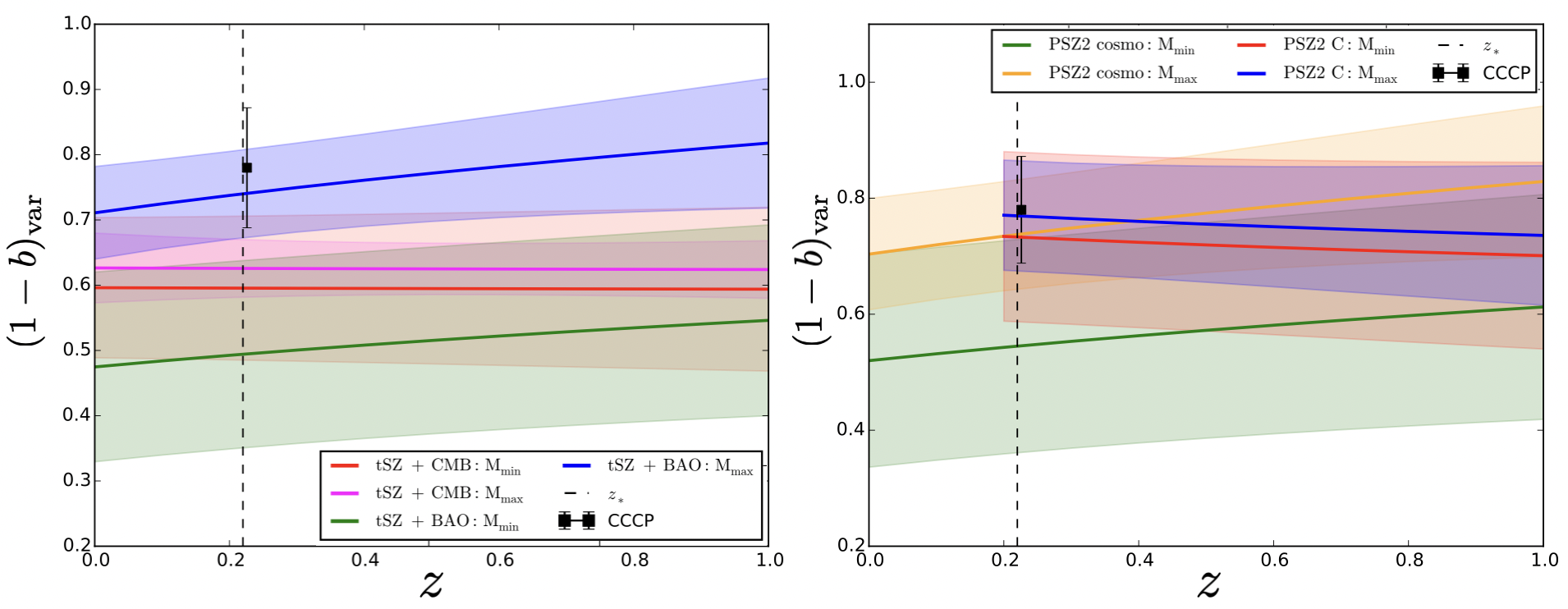}
\caption{Evolution of the mass bias with respect to redshift, at fixed values of mass, for different dataset combination. We compare results for the full Planck cosmological sample (left panel), with results obtained when selecting clusters with $z\ge 0.2$ (right panel). These plots are taken from \cite{Salvati:2019zdp}.}
\label{fig:p2_1}       
\end{figure}

\noindent We consider the tSZ probes defined above ($\text{NC}^{\text{tSZ}}$ + $C_{\ell}^{\text{tSZ}}$) and compare with results obtained when adding also CMB primary anisotropies. The full analysis is presented in \cite{Salvati:2019zdp}, we discuss here the main results.
On the one hand, when focusing on the tSZ+CMB dataset combination, we find that CMB data are driving the constraining power, not allowing for any evidence of mass and redshift evolution for the mass bias, and therefore finding consistent results with the ones discussed in section~\ref{sec:comb_tSZ}.
On the other hand, when considering tSZ probes alone, we find a mild evidence for a redshift evolution of the mass bias, with $\beta_b = 0.24 \pm 0.2$. We show this evolution for the mass bias in figure~\ref{fig:p2_1}. In the left panel, the green and blue curves represent the mass bias for fixed minimum and maximum values of mass respectively, $M_{\text{min}} = 0.83 \cdot 10^{14} M_{\odot}$ and $M_{\text{max}} = 14.69 \cdot 10^{14} M_{\odot}$. We find therefore a trend for an increasing mass bias with redshift.
However, this result is the opposite of what we would expect: as clusters evolve with cosmic time, we expect them to reach the HE condition towards $z=0$, reaching therefore $(1-b) \simeq 1$.  

In order to further understand our findings, we perform different consistency tests. We refer again the reader to the full analysis in \cite{Salvati:2019zdp} and report here the main results. We select a subsample of the full Planck cosmological sample, considering only clusters with $z \ge 0.2$. We choose this lower limit in redshift to be consistent with the range used for the CCCP calibration and to highlight the behaviour of Planck sample when it starts to show a mass-limited selection. We show the new results in the right panel of figure~\ref{fig:p2_1}. The red and blue curves represent the evolution of the mass bias for fixed $M_{\text{min}}$ and $M_{\text{max}}$ respectively, pointing to an opposite trend with respect to the previous results.

We stress therefore that these results are completely dominated by the adopted sample selection. This result is confirmed when we compare similar analysis performed in recent years (see a list of references in \cite{Salvati:2019zdp}): a possibile mass and redshift evolution for the mass bias produces different results depending on the selected cluster subsample.

We conclude with a general discussion for the results presented up to now: 
the mass bias parameter is introduced into the analysis to generally take into account the incomplete HE assumption. In practice, it may encompass different sources of uncertainties entering into the analysis, such as the model assumed for the cluster detection 
or the impact of the detection noise. Our results can also point to a general need of a more complete description for the scaling relations, improving the modelling of the interplay between astrophysics and cosmology in cluster formation and evolution and moving on from the simplistic assumptions of self-similar evolution and hydrostatic equilibrium.

We further improve this analysis, by combining Planck cluster sample with the SPT cluster observations \cite{SPT:2014wbo,SPT:2018njh}. We exploit SPT cluster cosmological constraining power to provide an independent calibration for Planck scaling relations. Among other results, we find a mass and redshift evolution for the mass bias that is consistent with the above analysis. This analysis is developed within the SPT collaboration and will be presented in \cite{Salvati_prep}.

\section{Halo mass function calibration}

Next generation experiments will provide samples of thousands of clusters suitable for the cosmological analysis. We need therefore to quantify the impact on cosmological constraints of the other assumptions we perform in the analysis, apart from the mass calibration. We focus here on the impact of the halo mass function, with a forecasts analysis.

When modelling the cluster number counts, for the mass function we adopt fitting formulas calibrated from numerical simulations.
In recent years, many analysis have provided different formulations and calibrations that can be used in the cosmological analysis (see e.g. the discussion in \cite{Monaco:2016pys} and references therein). It has been shown that the mass function calibration can impact the final results on cosmological parameters up to $\sim 10\%$, see e.g. the discussion in \cite{Paranjape:2014lga,Bocquet:2015pva}. Indeed, fitting formulas obtained from different numerical simulations may reflect different initial conditions and assumptions performed during the simulations (e.g. assumed initial cosmology, definition of the cluster mass and detection, resolution of the simulation).

In our investigation, we compare two different formulations for the mass function, both widely used in the cosmological community: \cite{Tinker:2008ff} (hereafter T08) and \cite{Despali:2015yla} (hereafter D16). We choose these two formulations since they represent two main approaches in evaluating the mass function (see e.g. discussion in \cite{Sakr:2018new}). We forecast results for three future surveys: the Euclid satellite \cite{EUCLID:2011zbd}, the LSST - Vera Rubin telescope \cite{LSSTScience:2009jmu} and the WFIRST - Nancy Grace Roman telescope \cite{Spergel:2015sza}. The full analysis is discussed in \cite{Salvati:2020exw}, we report here the main results.

\begin{figure}
\centering
\sidecaption
\includegraphics[width=4.5cm,clip]{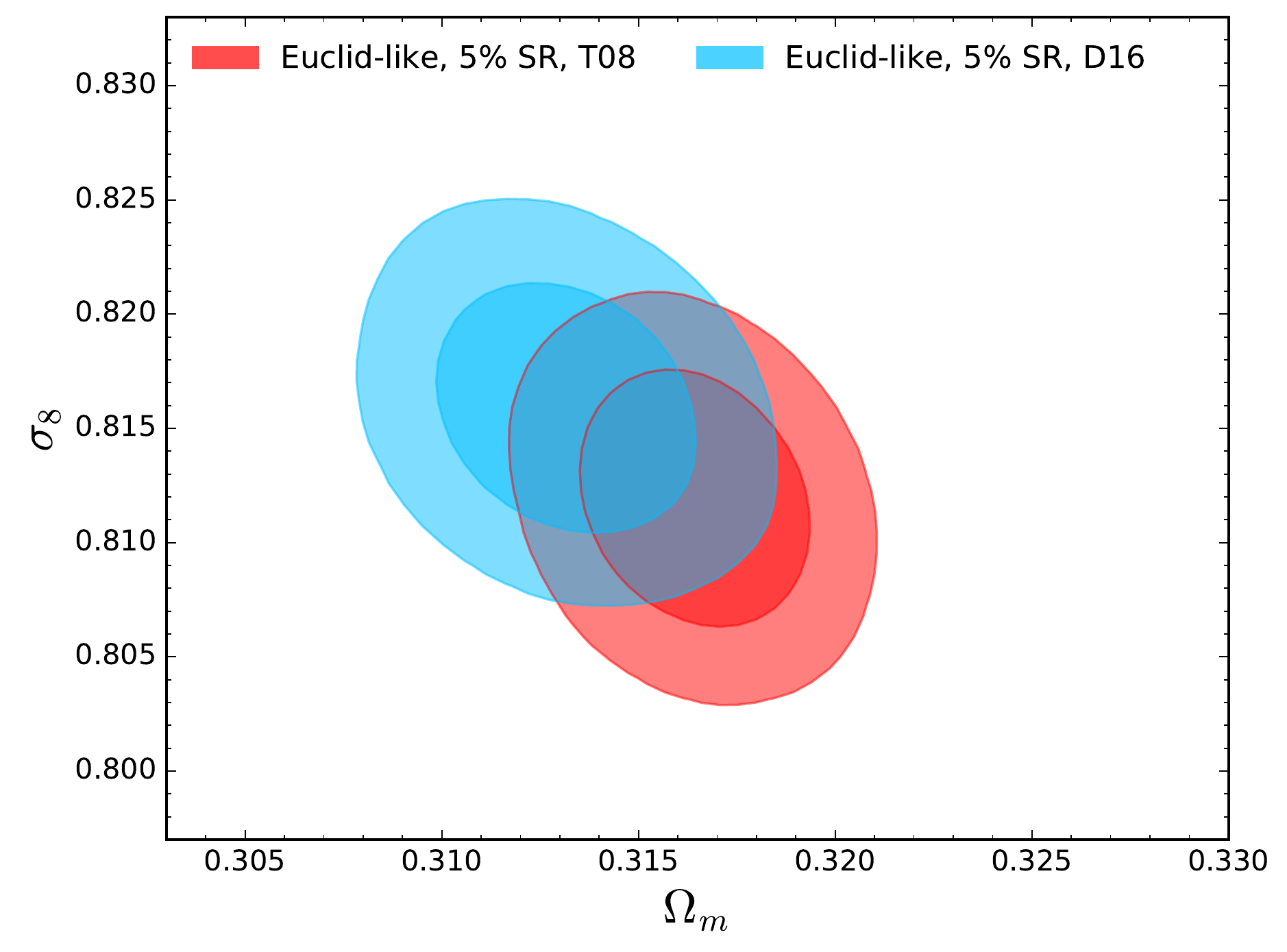} \, \includegraphics[width=4.5cm,clip]{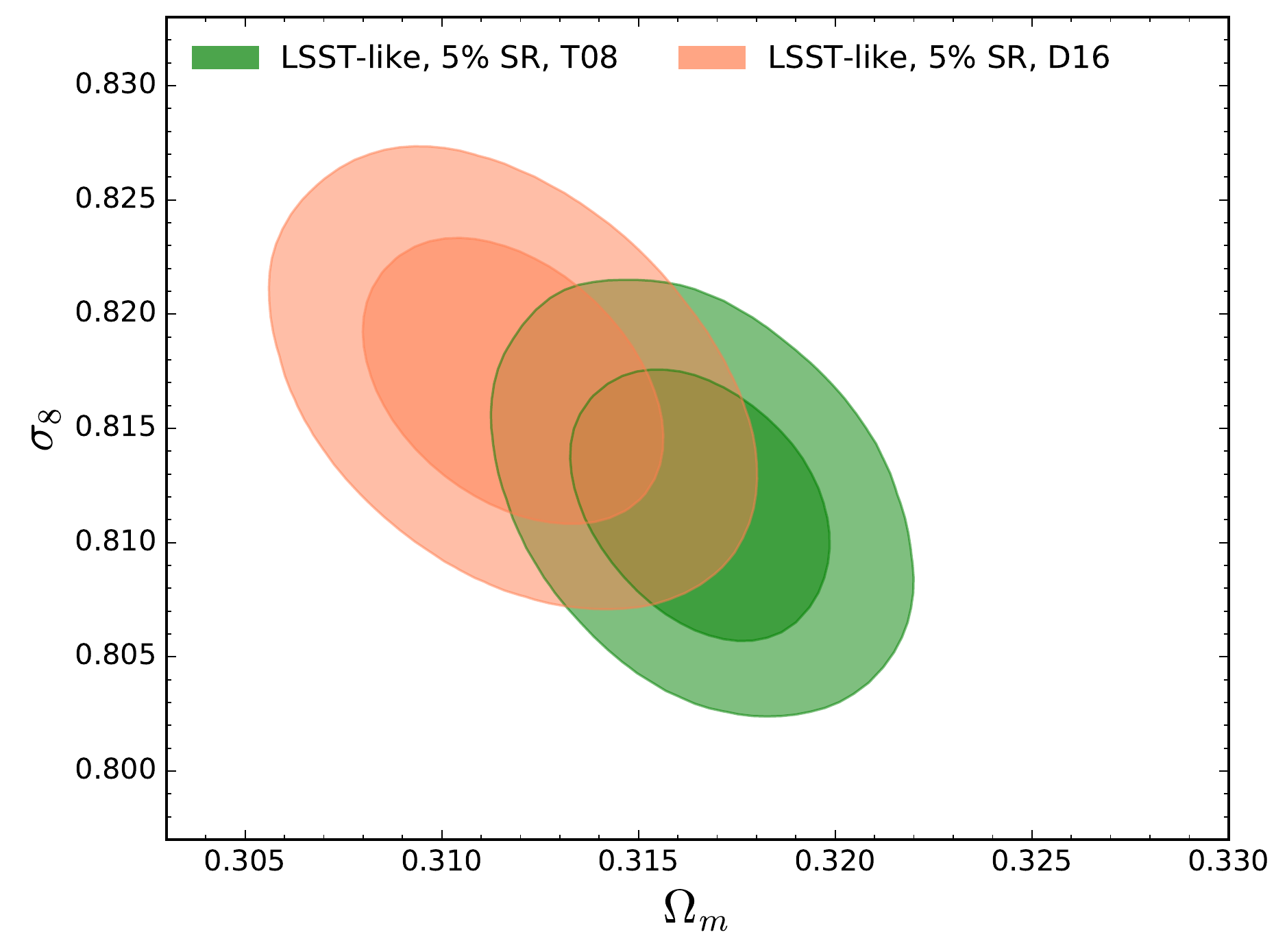} \, \includegraphics[width=4.5cm,clip]{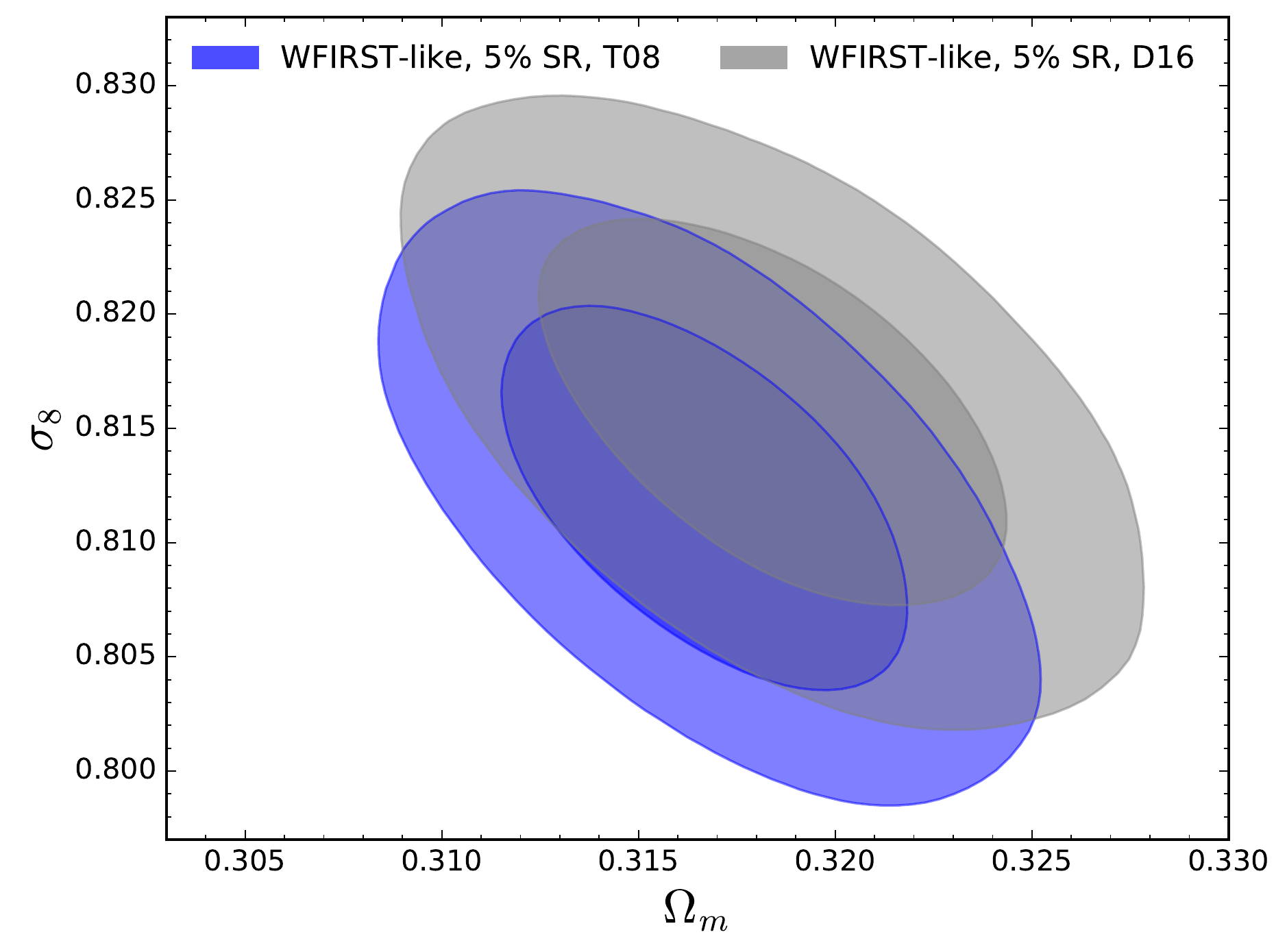}
\caption{Two-dimensional probability distributions in $(\Omega_m\sigma_8)$ parameter plane for the Euclid-like, LSST-like and WFIRST-like experiments. These plots are taken from \cite{Salvati:2020exw}.}
\label{fig:p3_1}       
\end{figure}

Assuming a $5\%$ accuracy on the calibration of the scaling relations, we compare the results we obtain on the cosmological parameters when adopting the two mass function formulations. We show the results in figure~\ref{fig:p3_1}, for the $\Lambda$CDM scenario. The two mass function evaluations produce different constraints, especially for the Euclid-like and LSST-like experiments, which are expected to produce larger cluster catalogs (due to the larger observed sky area). As discussed in \cite{Despali:2015yla}, the two mass functions show consistent behaviour in the intermediate mass and redshift range, while D16 predicts more cluster at high redshift. This difference in the predicted number counts seems to be the cause for the discrepancy we find in the cosmological parameters. In our analysis we adopt an intrinsic scatter for the scaling relations with a redshift dependence, following \cite{Sartoris:2015aga}:
\begin{equation}\label{eq:sigma_sr}
\sigma^2 _{\ln M}(z) = \sigma_{\ln M,0}^2 -1 + (1+z)^{\beta} \, .
\end{equation}

\noindent When adopting the two mass functions, for the redshift evolution of the scatter, we find $\beta = 0.1250 \pm 0.0025$ for T08 and $\beta = 0.1056 \pm 0.0025$ for D16, for the Euclid-like experiment (for which we have the larger impact), showing therefore a $\sim 8 \, \sigma$ discrepancy, which compensates for the high-redshift difference between the two mass functions. To further understand the impact of this different redshift evolution, we consider a non-constant EoS for dark energy, following the parametrisation $w = w_0 +(1-a)w_a$ \cite{Chevallier:2000qy,Linder:2002et}. Still focusing on the Euclid-like experiment, for the two mass functions we find $w_0 = -1.000 \pm 0.010$ and $w_a = 0.35^{+0.32}_{-0.13}$ for T08, $w_0 = -1.095 \pm 0.012$ and $w_a = 0.13^{+0.51}_{-0.26}$ for D16, while having consistent results for the $\beta$ parameter. In this case, the different redshift evolution between the two mass functions produces the same $\sim 8 \, \sigma$ discrepancy on the $w_0$ parameter: a different EoS for dark energy implies a different expansion rate for the background, consistent with the different redshift evolution for the mass functions.

From this analysis, the halo mass function emerges as a non-negligible source of systematic uncertainties. It is therefore necessary to find a ``universal" calibration, independent on the adopted simulation.

\section{Conclusions}
Galaxy clusters are emerging as a powerful cosmological probe, useful to describe the latest evolution of the large scale structure. The accuracy and precision of the cosmological constraints inferred from cluster analysis are dominated by theoretical and observational modelling. In order to improve the analysis, we would need further developments in hydrodynamical simulations and precise description of the astrophysical processes affecting the cosmological evolution of clusters. The exquisite quality of future cluster observations (such as NIKA2 follow-up program) will help in characterising the cluster properties.

\end{document}